
\input phyzzx
\unnumberedchapters
\vsize 23.7 true cm
\hsize 16.5 true cm
\baselineskip=15.7 true pt

\font\twelvegoth=eufm10 at 12pt
\newfam\gothfam
\textfont\gothfam=\twelvegoth
\def\goth{\fam\gothfam\twelvegoth}

\font\twelveBB=msym10 at 12pt
\newfam\BBfam
\textfont\BBfam=\twelveBB
\def\BB{\fam\BBfam\twelveBB}

\tolerance=3000
\def\piedepagina{
 \footline={\ifnum\pageno>1 \hss\tenrm\folio\hss
            \else \hfil \fi}}
\mathsurround=2pt

\piedepagina

%

\def\D{\cal D}

%
%
\def\mn{{\mu\nu}}
\def\a{\alpha}

\def\m{\mu}
\def\n{\nu}
\def\r{\rho}

\def\d{\delta}
\def\dd{\Delta}

\def\ee{\epsilon}

\def\gm{\Gamma}
\def\gmb{\bar{\gm}}

%
%
\def\tp{\tilde{p}}
\def\hp{\hat{p}}

\def\ird{\underline{\omega}}

\def\WW{{\scriptscriptstyle R}}
\def\VV{\scriptscriptstyle V}
\def\RR{{\rm I\!\!\, R}}

\def\RR{{\rm I\!\!\, R}}
\def\cv{c_{\VV}}

\def\yy{{\cv\over k}}
\def\idx{\int d^3\!x}

%
%

%
%

%
%
\def\dG{{{\d \phantom{G}}\over {\d G_{\m}}}}
\def\dgmbG{{{\d\gmb}\over {\d G_{\m}}}}

%

%
\def\da{{{\d}\phantom{A}\over {\d A^{\m }}}}
\def\dgmba{{{\d\gmb}\over {\d A^{\m }}}}

%
%

%
%
\def\dc{{{\d}\over{\d c}}}
\def\dgmbc{{{\d\gmb}\over{\d c}}}

%
%
\def\dH{{{\d}\phantom c\over{\d H}}}
\def\dgmbH{{{\d\gmb}\over{\d H}}}

\def\Scs{S_{\rm CS}[A\,]}
\def\scv{{\rm sign}(k)\cv}
\def\ll{\Lambda}
\def\sk{ {4\pi\over k}\, }
\def\frac #1#2{{#1\over #2}}
\def\set#1{\left\{ #1 \right\} }            


{\rightline{LPTHE 92-42}}
\vskip .7true cm
{\centerline{\fourteenbf Regularization and Renormalization of
Chern-Simons Theory \foot{Talk presented by G. Giavarini at the
NATO AWR on ``Low dimensional Topology and Quantum Field Theory'',
6-13 September 1992, Cambridge (UK).} }
\vskip 0.9 true cm

\centerline{G.~Giavarini,$^{\rm a)}$  C. P.~ Martin$^{\rm b)}$
and
F.~ Ruiz Ruiz$^{\rm c)}$ }
\vskip 0.7 true cm
{\tenpoint
\leftline{a) L.P.T.H.E.,
Universit\'e Paris VI-VII,
2 Place Jussieu, 75251 Paris Cedex 05, France.}
\vskip 0.2 true cm
\leftline{b)
 Department of Mathematics and Statistics,
     University of Guelph,
         Guelph, Ontario N1G 2W1, Canada.}
\vskip 0.2 true cm
\leftline{c)
 Niels Bohr Institute, University of Copenhagen,
Blegdamsvej 17, DK-2100 Copenhagen \O, Denmark.}
}

\vskip .8 true cm

The topological field theory most familiar both to physicists and to
mathematicians is surely Chern-Simons theory.  The classical
Chern-Simons euclidean action, for a principal $G$-bundle $P$ over an
oriented three manifold $M$, is given by $$
\Scs =
-{ik \over 4\pi} \int_M \Tr \left(A\wedge {\rm d} A +
{ 2 \over 3} A\wedge
A\wedge A \right),
\eqn\coofree
$$ where $A$ is the gauge connection on $P\to M$ and $\Tr$ represents
a $G$-invariant bilinear form on $\goth g$, the Lie algebra of $G$. In
the following we shall assume $G=SU(N)$ and $A$ taking values
 in the fundamental
representation of ${\goth su}(N)$, with the (antihermitian) generators
$T^a$ normalized so that $\Tr (T^a T^b)={1\over 2}\,\d^{ab}.$ The field
theory with action \coofree\ is ``topological'' in the sense
that it does not depend on the metric chosen on $M$ and general
covariance is thus manifest.

\REF\Iengo{For a recent review see for example: \nextline R. Iengo and K.
Lechner, Phys. Rep. {\bf 213} No. 4 (1992) 179.}
The action \coofree\ is invariant under the infinitesimal gauge
transformation $A\to A+D\eta$, where $\eta$ is a ${\goth su}(N)$
valued function on $M$ and $D=d+A$ is the covariant derivative with respect
to  $A$.
The field theory described by $\Scs$ is then
particularly appealing since it provides an alternative to the usual
Yang-Mills action for describing gauge fields in three dimensions.
It is by now a well established fact that matter fields coupled to
a Chern-Simons gauge field give rise to particles with fractional
statistics (the so called anyons) that might be relevant for the
description of phenomena such as the fractional quantum Hall effect or
high $T_c$ superconductivity [\Iengo].
In the following however, our main concern will be  the pure
gauge theory with action given by \coofree.

As usual in field theory, one is interested in the computation of the
vacuum expectation value of observables.  It is clear that gauge
invariant local operators are not necessarily generally covariant and
consequently they are not appropriate for the case at hand.  One must
then look for  non-local, metric independent and gauge invariant
objects.  Physicists working on more standard gauge theories (like the
old good QCD in 4 dimensions) have been for
a long time dealing with the
computation of observables having the above properties. They called
these observables
``Wilson loops'', but mathematicians seem to prefer the name of
``holonomies'' of the connection $A$. Given a closed curve $C$ in $M$,
a Wilson loop is  defined as
$$ {\cal W}(C)=\Tr P \exp\oint_C A ~~.$$
The expectation value  of a
collection $\set {C_i}$ of Wilson loops  is
given  by the Feynman path integral
$$
Z(M;C_i)=\int {\D} A\> \prod_i
{\cal W}(C_i) \,{\rm e}^{-\Scs} ~~.
\eqn\naive
$$
In the simple case in which no Wilson
loop appears in \naive\
 one  gets   the partition function $Z(M)$ of the
theory.
\REF\Witten{E. Witten, Commun. Math. Phys. {\bf 121} (1989) 351.}
\REF\Jones{V.F.R. Jones, Ann. Math. {\bf 126} (1987) 335.}
\REF\Freed{D.S. Freed and R.E. Gompf, Phys Rev. Lett. {\bf 66} (1991) 1255.}

Owing to the intrinsecally topological character of $\Scs$ one expects
the functors $Z(M;C_i)$ to give back topological invariants of $M$ and
$\set{C_i}$.  To make
sure this is indeed the case we might content ourselves with the study
of the
formal properties of the path integral \naive.  However, if some
explicit result has to be obtained, one cannot avoid to give concrete
meaning to eq.\naive. This in practice signifies that we must quantize
Chern-Simons theory (CST).  Taking $M=S^3$,  it was
shown in ref. [\Witten],
by using input results coming from conformal quantum field
theory, that the expectation values of Wilson loops satisfy the same skein
relation as the Jones polynomials [\Jones] and can be identified with
the latter.  More precisely, $Z(S^3;C)/Z(S^3)$
with the standard framing is the Jones polynomial for
$C$ in the variable $q=\exp \{ 2\pi i/(k+\scv) \}$, where $k$ is the
classical parameter appearing in \coofree\ and $\cv$ is the second
Casimir operator in the adjoint representation of $SU(N)$. With our
normalization choice for the group generators, $\cv$ is simply $N$.
In this way CST provides  a three dimensional framework
for computing knot invariants. The partition function itself
is a topological invariant of $M$, as  explicitly checked
in ref.  [\Freed].

The fact that the observables of CST are  functions of the combination
$k+\cv$, rather than  $k$, has a clear quantum mechanical
interpretation. It relies on the precise correspondence between CST on
$M$ and the Wess-Zumino-Witten model on $\partial M$, as first displayed
in [\Witten]. Indeed, $k$ can be identified with the level of the
affine $SU(N)$ current algebra on $\partial M$.  Then
the Sugawara form of the energy-momentum tensor implies the shift
$$k\to k+\scv ~~.\eqn\shift $$
We shall discuss in a moment the reasons leading to
quantization conditions for the parameter $k$ in CST which, just like the
central charge in the representation theory of affine algebras, must
take integer values.

\REF\Guadagnini{E. Guadagnini, M. Martellini and M. Mintchev,
 Nucl. Phys. {\bf B330} (1990) 575.}

One may think of getting a clue to
the understanding of the shift \shift\  without
resorting to the aid of conformal field theory.
In a semiclassical quantization of CST, eq.\shift\
can be recovered as the radiative correction to the
parameter $k$ at order $\hbar$ [\Witten].  {}From this simple remark,
it should appear clear  the relevance of perturbation theory to acheive
a   complete understanding of the quantum theory defined by
\coofree. Perturbation theory,
moreover, provides a very direct tool for obtaining intrinsically three
dimensional integral representations of knot and link invariants that
generalize  Gauss' formula for the linking number of two closed curves, as
illustrated in ref. [\Guadagnini].

\REF\GuadagniniPL{E. Guadagnini, M. Martellini and M. Mintchev,
Phys. Lett. {\bf 224} (1989) 489.}
\REF\Alvarez{L. Alvarez-Gaum\'e,
J.M.F. Labastida and A.V. Ramallo, Nucl. Phys. {\bf B334} (1990) 103.}
\REF\GMR{G. Giavarini, C.P. Martin and F. Ruiz Ruiz,
 Nucl. Phys. {\bf B381} (1992) 222. }
\REF\Asorey{M. Asorey and F. Falceto,  Phys. Lett. {\bf 241} (1990) 31;
 Int. J. Mod. Phys. {\bf A7} No.2 (1992) 235.}
\REF\review{G. Giavarini, F. Ruiz Ruiz, C.P. Martin, {\it Perturbative
quantization of Chern-Simons theory}, preprint.}

 Quite a number of papers dealing with the issue of
perturbative quantization of CST have already appeared
[\GuadagniniPL-\review].  This has led to some controversy about the exact
meaning of the shift \shift\ in the perturbative framework.  Our main
concern here will be to provide a sound perturbative setting for the
quantization of CST. A careful analysis of some of the features of the
perturbative approach  will result in a proper understanding of eq.
\shift.
Our study  passes through the determination of the effective
action $\Gamma[A\,]$, the quantum analogue of the classical $\Scs$. The
effective action is the generating functional of the 1PI Green
functions and therefore the starting point for the computation of any
other quantum observable.
\REF\Deser{S. Deser, R. Jackiw and S. Templeton,
Ann. Phys. {\bf 140} (1982) 372.}

 We have already mentioned the invariance of $\Scs$ under
infinitesimal gauge transformations.  Acting with a {\it
finite}  gauge transformation $h$, $\Scs$ is transformed into [\Deser]
(apart from surface terms)
to $S_{\rm CS}[^h A\,]=\Scs + 2\pi k \, {\rm w}(h)$.
The quantity ${\rm w}(h)$ is an integer, being the winding number
of $h$.
In order to have a sigle valued partition function, $k$ must
obey the quantization condition $k\in {\BB Z}$.
Notice that this
 is a non-perturbative requirement, however, since finite gauge
transformations lie beyond the perturbative regime.  Furthermore, at
the quantum level, the same requirement of monodromy should not be
imposed on the classical $k$ but rather on its quantum or renormalized
counterpart. We shall come back to this point later on.

To  quantize perturbatively CST one must first fix a gauge.
We shall work in the Landau gauge, which is known to be free of
infrared divergences. With the standard Faddev-Popov construction this
amounts to adding to $\Scs$ the term
$$
S_{\rm GF} =
2 \int_M \, d^3\!x ~ \sqrt g ~\Tr \left[ ( J^{\m} -
\partial^\m \bar{c} ) D_\m c -b\partial_\m A^\m - \frac 12 \,
 H [ c, c] \right] ~ .
$$
As customary, $b$ denotes the Lagrange multiplier for the gauge
condition $\partial_\m (\sqrt g A^{\m})=0,
$ $c$ and $\bar{c}$ are Faddeev-Popov
ghosts and $J$ and $H$ are
auxiliary fields introduced for later convenience.
The  relevant aspect of $S_{\rm GF}$ is that it necessarily picks
a metric $g$ on $M$.
Thus the  resulting gauge-fixed action $S=S_{\rm CS}+S_{\rm GF}$
is no longer  gauge invariant nor metric independent.
The action is, however, BRS invariant. As $S_{\rm GF}$ is a pure BRS
variation, it is not observable. Therefore, at least at the classical
level, not only  gauge invariance but also the topological
character of the theory is recovered. One might then wonder if the
same result holds true for the quantized theory.
\REF\Blasi{ A. Blasi and R. Collina, Nucl. Phys. {\bf B345} (1990) 472.}
\REF\Delduc{F. Delduc, C, Lucchesi, O. Piguet and S.P. Sorella,
Nucl. Phys. {\bf B346} (1990) 513.}

To carry out our quantization
program we choose to work on $M={\RR}^3$ endowed with the flat
metric $g_\mn=\delta_\mn$.
Although by naive power counting the theory appears renormalizable, it
is in fact ultraviolet (UV) finite [\Blasi,\Delduc], that is, the beta
function and the anomalous dimensions of the fields vanish to all
orders in perturbation theory.  This result should not be very
surprising since CST on $ S^3$, being topological, has no physical
local excitations.
It was subsequently
questioned if  UV finiteness of CST implied the vanishing of radiative
corrections to the parameter $k$, something which would be
in disagreement with the well-established
non-perturbative result \shift. We shall see that this is not going to
be the case.


UV perturbative finiteness of Green functions does not imply that the
corresponding Feynman diagrams should also be finite.  On the
contrary, since power counting predicts divergences for individual
diagrams, the expected scenario is a cancellation of the divergences
order by order in perturbation theory when summing over all diagrams
contributing to a given Green function.  Although divergences cancel
out in the final answer, for practical computational reasons the
divergent integrals must be made mathematically manageable by means of
a suitable regulator. In this respect CST does not differ from
ordinary renormalizable theories.

Let us denote with $\ll$ the regulator
(or set of regulators) needed to regularize CST at any
perturbative order.  Since the theory is UV finite, the limit
$\ll\to\infty$ in which the regulator is removed defines a
``minimal'' renormalization scheme where renormalized quantities equal
bare ones.  In this scheme, that we call renormalized=bare,
 the renormalized effective action
$\Gamma$ and  Green functions are defined as
$$
\gm\, =\,
      \lim_{\ll\to \infty}\,
             \gm _\ll ~ , \qquad
 G(p_1,\ldots ,p_E)= \lim_{\ll\to\infty}  \>
G_\ll (p_1,\ldots ,p_E) ~~,
\eqn\Llimit
$$
where $\gm _\ll$ and $G_\ll$ are the corresponding regularized
quantities.
The value of the Green functions so obtained depends in principle on the
particular regulatization employed and so does the value of the observables
of the theory. However, we shall see that all the known regularization methods
satisfying  certain invariance requirements lead  to  observable
one-loop radiative corrections that reproduce the non-perturbative
results.
Thus, these invariance requirements, along
with the renormalization scheme above,
provide  a ``natural'' parametrization of the perturbative theory.


The BRS symmetry of the gauge fixed action is what is left of the
original symmetries of the theory.  Classically, the
cohomology of the BRS operator guarantees that, when computing
observables, the original symmetries still hold true. It is then clear
that if we set some hope on obtaining the same picture at the quantum
level, we cannot but rely on the BRS symmetry, which is known to be
non anomalous in this case [\Blasi,\Delduc]. We then say that BRS is {\it
fundamental}, meaning with this that it must be enforced at the
quantum level in order to make sense out
 of the quantum theory.  It will
appear that for CST in the Landau gauge, once BRS symmetry has been
implemented on the renormalized theory, the parameter $k$ occurs to
be	 the only actual free parameter.  The ``natural''
parametrization choice is then in terms of the bare (or classical)
$k$.  This does not specify completely a renormalization scheme
because  finite renormalizations of the fields are
still allowed.  However, since the value of the observables is
unchanged under finite wave function rescalings, the quantum
theory  is unambigously defined. Actually, we shall make
recourse to this freedom of rescaling the fields
when comparing Green functions obtained with
different regulators.
The simplest way to make sure that we get a BRS invariant quantum
theory is to start with a BRS invariant regulator so that the effective
action we get from \Llimit\ in the scheme renormalized=bare satisfies
automatically the BRS identity. It will turn out that the observables
as functions of the bare $k$ are the same functions of $k+\cv$ for
all BRS invariant regularization schemes. Obviously, the agreement
between  perturbative and non-perturbative results in this case
originates from the thorough gauge invariance of both quantization
methods. It is worth mentioning that with explicitly BRS breaking
regularizations, such as the one in ref. [\GuadagniniPL], the shift
\shift\ is not observed as consequence of the loss of BRS invariance
at the regularized level, despite the latter is restored when the
regulator is removed.

We then move on to the
computation of the most general BRS invariant effective action.  It
is convenient to introduce the functional
$$
\gmb\,=\,\gm + 2\idx\,\, \Tr( b \>\partial_\m A^\m)
$$
which, owing to the Landau gauge condition and the antighost equation,
is idependent of
$b$ and depends  on  $J_{\mu}$ and ${\bar c}$ only through the
combination $G_{\m}=J_{\m} -\partial_{\m} {\bar c} $ [\GMR].
The BRS identity for $\gmb$ takes then the  form
$$
\idx \Tr \left( \dgmba\,\dgmbG + \dgmbc \, \dgmbH \right)
        \,=\,0 \,\, .
\eqn\brseq
$$
Inserting in eq.\brseq\ the loop expansion
$\gmb = \sum_{n=0}^{\infty} \hbar^n \gmb_n$, we get a tower of equations
that must be satisfied order by order in $\hbar$. At first order we
have
$$
\Delta \gmb_1 =0 ~~,
\eqn\first
$$
where $\Delta$ is  the Slavonv-Taylor operator
\def\dder#1#2{ {\delta #1 \over \delta #2}}
$$
\dd \,=\, \idx \Tr \left[ \dder{\gmb_0}{A^{\m }}\, \dG
+ \dder{\gmb_0}{G_{\m}}\, \da
          + \dder{\gmb_0}{c} \, \dH + \dder{\gmb_0}{H}
\, \dc \right]\,\, .
$$
The operator $\Delta$ is nihilpotent and is the quantum generalization
of the classical BRS operator.

Eq. \first\ is formally identical to the usual stability equation for
BRS. However, the cohomological problem to be solved here is by far
more difficult, since $\gmb_1$ might contain not only local but also
non-local contributions.
Under the additional hypothesis that contributions involving fewer
than four fields be purely local, the local and non-local sectors can
be proven to decouple in eq.\first\ and we can easily solve  for the
local part [\GMR].  This locality requirement is justified a
posteriory by the explicit computation of the Green functions
involving up to three external legs. Observe that the latter are the
only Green functions that by power counting need to be regularized.
Consequently, the local
part of $\gmb_1$ encodes all the arbitrariness of the regularization
scheme used.  Of course, terms involving four or more fields are
necessarely non-local, if not zero, for dimensional reasons and, being
finite, do not depend on the regularization.
For the local part of $\gmb_1$ we then get [\GMR]
$$
\eqalign{
W(\a_1,\beta_1,\gamma_1) \,= & -{ik \over 4\pi} \idx \,\,
       \Tr \left[ \left( \a_1+2\,\beta_1 \right)\,
             A \wedge {\rm d} A +
       {2\over {3}}\, \left( \a_1+ 3\,\beta_1 \right)\,
             A\wedge A\wedge A \right] \cr
       & ~~~- 2\idx\, \Tr \left[\,  \beta_1\, G_{\m} \,\partial^{\m}c
             - \gamma_1\, G_{\m}{\left( D^{\m}c \right)}
             + {1\over 2}\,\gamma_1\, H [c,c] \right],
\cr}
$$
where $\a_1$, $\beta_1$ and $\gamma_1$ are regularization dependent (finite)
coefficients. They correspond to the freedom of renormalizing  $k$ and
the fields.  The (local part of the) one loop
effective action can then be recast in the form
$$
 \gm^{\rm loc} \,=\,
(1+\a_1 ) S_{CS}[A\,] \>
- \,2\idx\,\, \Tr (b \, \partial_\m A^\m)
\,\,+\,\,\Delta \,X(\beta_1,\gamma_1) ~ .
\eqn\beautiful
$$
Here $X(\beta_1,\gamma_1)=\,4\idx \Tr \left[ \beta_1 G^{\m}
A_{\m} -(1+\gamma_1) H c \right]$.
The effective action receives  two different kinds of
contributions. One (corresponding to $\a_1$) is gauge
invariant, metric independent and provides  a monodromy parameter $(1+\a_1)k$.
The other is the  term $\dd X$ which, being
cohomologically trivial, does not contribute to to the expectation
values of  Wilson loops
[\review].   These properties can be made manifest by
the wave function renormalization $\Phi_\WW=Z_\Phi\Phi$
($\Phi=A,b,c,H,G$), with
$Z_A\,=\,Z_G^{-1}\,=\,Z_b^{-1}=1+\beta_1$,
$Z_c\,=\,Z_H^{-1}=1+\gamma_1$, so that $\gm^{\rm loc}$ takes the
form
$$
\gm^{\rm loc}\,=\,
        (1+\a_1)\, S_{\rm CS}[A_\WW ]
+S_{\rm GF}[\Phi_\WW] ~~.
\eqn\soneloop
$$
Since $\a_1$, the only observable one-loop correction,
could still depend on the regularization, one
 would  conclude that the emerging quantum theory is not
unique.   As anticipated, however, all BRS invariant
regulators used so far for CST yield the same value for $\a_1$.  In
the folowing table we collect the values (in units of $\cv /k$) of the
one-loop parameters for the following BRS invariant regularization
schemes: large mass limit of dimensionally
regularized topologically massive Yang-Mills (TMYM) theory [\GMR],
$\eta$-function regularization [\Witten], higher covariant derivatives
(HCD) and Pauli-Villars fields [\Alvarez,\review] and geometric
regularization [\Asorey].
\foot{The values given here for HCD+Pauli-Villars are
those computed in [\review] rather than those in [\Alvarez], where
strictly speaking
only Pauli-Villars fields and no HCD are used. In [\Alvarez] one
gets $\a_1=1$, $\beta_1=\gamma_1=0$.}
$$
\vbox{
\offinterlineskip
\def\tablerule{\noalign{\hrule}}
\def\homit{height3pt&\omit&&\omit&&\omit&&\omit&\cr}
\hrule
\halign{&\vrule#&
\strut\quad\hfil#\hfil\quad\cr
\homit
& Regularization Method \hfil&& $\a_1$ && $\beta_1$ && $\gamma_1$ &\cr
\homit
\tablerule
\homit
& Large $m$ of TMYM + dimensional reg. && 1 && 2/3 && 0 &\cr
\tablerule
\homit
& $\eta$-function regularization && 1 && 0 && 0 &\cr
\tablerule
\homit
&HCD + Pauli-Villars   && 1 && 2/9 && 0 &\cr
\tablerule
\homit
& Geometric regularization && 1 && $4/(3\pi)\> I_n$ && $\ast$ &\cr
\homit }
\hrule
}
$$
Geometric quantization makes use of ghost generations different from
the standard Fad\-deev-Popov ones, so only the pure gauge sector of
$\gm^{\rm loc}$ can be compared; the quantity $I_n$ is defined as
$I_n=\int_0^\infty dp\>\> {(1+p^2)^n \over 1+p^2(1+p^2)^{2n} }$, with
$n>1$ an arbitrary integer.
As it appears evident, different regularizations yield different
$\Gamma$'s but all of them
predict the same observable shift $k\to k+\cv$ and therefore the same
physical theory. This  is in agreement with the non-perturbative
result \shift\ and   leads to a self-consistent, single valued
quantum theory, since the monodromy parameter $k+\cv$ keeps beeing
an integer.

At this point the obvious question arises: what does it happen at
higher perturbative orders? Now that we obtained the solution
\beautiful\ for the one loop effective action, we can proceed and
solve the BRS identity at order $\hbar^2$.  If $\gmb_2$ has the same
locality properties as $\gmb_1$, its local part is found to be [\GMR]
$$
\gmb_2^{\rm loc}=W(\a_2,\beta_2,\gamma_2)
+{ik\over 4\pi} \idx \, \beta_1(\a_1+3\beta_1-\gamma_1)
\, \Tr(  A \wedge {\rm d} A ) ~~~
$$
so that the (local part of) the two-loop effective action reads
$$
\Gamma^{\rm loc}=
\Gamma_0+W(\a_1+\a_2,\beta_1+\beta_2,\gamma_1+\gamma_2)
+{ik\over 4\pi} \idx \, \beta_1(\a_1+3\beta_1-\gamma_1)
\, \Tr(  A \wedge {\rm d} A ) ~~~
\eqn\awkward
$$
The three new parameters $\a_2$, $\beta_2$ and $\gamma_2$ are second
order and regularization scheme dependent.  The apparently awkward
expression \awkward\ takes a simple and familiar form if we subtract
the one loop contributions corresponding to $\Delta
X(\beta_1,\gamma_1)$ entering the two-loop diagrams through one loop
subdiagrams. This is accomplished by means of the same wave function
renormalization leading to eq.
\beautiful. In terms of the renormalized fields we have
$$
\eqalign{
\gm_{\rm local}\,=\,
  & (1+\a_1+\tilde\a_2)\,\,S_{CS}\,[A_\WW] \,
      - \,2\idx\,\, \Tr( b_\WW \,\partial_\m A_\WW^\m) \,
+\,\dd_\WW X_\WW(\tilde \beta_2,\gamma_2) ~~, \cr}
$$
where the subsript $R$ indicates renormalized quantities and $
\tilde  \a_2=  \a_2 -6 (\beta_1)^2 -3  \a_1 \beta_1
+3 \beta_1 \gamma_1$, $\tilde \beta_2= \beta_2 +\beta_1^2 - \beta_1
\gamma_1$. Just like at one loop, BRS invariance implies that out of
three parameters, only one, $\tilde\a_2$, is gauge invariant hence
observable as a shift of $k$. Thus, from a perturbative viewpoint a
two-loop correction to $k$ would be allowed. Note that now
$\tilde\a_2$ is order $(\cv /k)^2$ and cannot possibly lead to an
integer shift for any integer choice of $k$.  Finding a non-zero
$\tilde\a_2$ would imply an incompatibility between the
non-perturbative request of single valuedness and  perturbative
quantization. The computation of $\tilde\a_2$ in a BRS invariant
regularization scheme is then particularly intriguing.
In ref. [\GMR] we showed how to compute the two-loop effective action
for the first BRS invariant regularization scheme listed in the table
above. In the remaining part of this note we shall briefy summarize
the ideas and the results of the method, warmly inviting the
interested reader to refer back to the original paper where more details can
be found.
\REF\Breitenlohner{P. Breitenlohner and D. Maison, Commun.
Math. Phys. {\bf 52} (1977) 11.}

Dimensional regularization   ensures BRS invariance and
algebraic consistency if we start with a $D$ dimensional
extended action $\Scs$ with the following prescription for the $\ee_{\mn\rho}$
symbol (showing up once $\Scs$ is written in coordinates).
The $D$-dimensional analogue of $\ee_{\mn\rho}$ is
defined as a completely antisymmetric object in its indices
satisfying the following identities [\Breitenlohner]:
$$
\ee_{\m_1\m_2\m_3}\ee_{\n_1\n_2\n_3}=
   \sum_{\pi\in {\cal S}_3} \, {\rm sign}(\pi)\prod_{i=1}^3 \,
         \tilde{g}_{\m_i\n_{\pi (i)}} \quad , \quad
\ee_{\m_1\m_2\m_3}\hat{g}^{\m_3\m_4}=0 \quad  .
\eqn\oeps
$$
Here $g_{\mn}=\tilde{g}_{\mn}\oplus\hat{g}_{\mn}$
is the euclidean metric in $D$ dimensions, with $\tilde{g}_{\mn}$
and $\hat{g}_{\mn}$ its 3- and $(D-3)\!$-dimensional projections
respectively.
Furthermore, given a $D\!$-dimensional vector $u^\m$ we define
$\hat{u}^{\m}=\hat{g}^{\mn}\,u_\n$ and
$\tilde{u}^{\m}=\tilde{g}^{\mn}\,u_\n.$ Notice that objects with a
hat vanish when $D\to3;$ they are called evanescent.

Unfortunately, pure CST theory with
the above prescription for the $\ee_{\m\r\n}$ symbol, has a
non-invertible kinetic term,  even  in a general $\a$-gauge,
because of the zero modes $z(p)=f(p)(\hat
p^2 \hat g_\mn -\hat p_\m \hat p_\n)$. This hinders a
perturbative analysis.  To get out of this empasse we can add to
$\Scs$ a Yang-Mills term
$$
S_{\rm YM}[A\,]={k\over 8\pi m} \idx \Tr (F_\mn F^\mn) ~~,
$$
where $F_\mn$ is the field strength of the gauge field $A_\m$.  The
theory defined by the action $S_m[A\,]=S_{\rm CS}[A\,]+S_{\rm
YM}[A\,]$ corresponds to so the called Topologically Massive Yang
Mills theory (TMYM), proposed in ref.  [\Deser] to provide a way
(alternative to the usual Higg mechanism) for making the gauge field
massive without losing gauge invariance. The parameter $m$ is the bare
topological mass of the gauge field and CST is recovered in the limit
of infinite mass gap. The TMYM theory, differently from CST, is
superrenormalizable and only one- and two-loop diagrams are divergent.
Therefore, from our viewpoint, $m$ can be envisioned as an UV
regulator.  The regularized CST is then defined in the limits
$m\to\infty$ and $D\to3$. These two limits do not commute and, to have
a consistent BRS invariant regularization prescription, one must take $D\to3$
first. Only in the case that the limit $D\to3$ does not give rise to
divergences, is one allowed to let $m\to\infty$. Therefore, if our
regularization prescription has to make sense, TMYM theory must be UV
finite.  That this is indeed the case has been shown in ref.  [\GMR].
\REF\Polychronakos{A.P. Polychronakos, Ann. of Phys. {\bf 203}, No.2,
(1990) 231.}
\REF\Klauder{J. Klauder and E. Onofri, Int. J. Mod. Phys. {\bf A4}
(1989) 15.}

The addition of a Yang-Mills term to the Chern-Simons action entails
several side effects.
{}From a non-perturbative, functional point of view, the wild
oscillatory behaviour of the path integral \naive\ is tamed by the
presence of the Yang-Mills contribution. Thus the TMYM path integral
is in fact a regularized version of Chern-Simons one.  The price paied
for this regulating effect is a total change in the structure of the
Hilbert space.  Indeed TMYM theory is not topological and the gauge
fields excitations are propagating.  In the limit of infinite mass
gap the propagating modes decouple from the
non-propagating ones,  and CST is reobtained.
What we are left with is thus the (degenerate)
ground state of TMYT which precisely corresponds to the zero-energy
Hilbert space
of CST [\Polychronakos]. This picture is reminiscent of  Klauder's
regularization prescription of the path integral for quantum
mechanical systems, which is known to be completely
equivalent to geometric quantization [\Klauder].

 The definition \oeps\ for
the $D\!$-dimensional $\ee_{\m\n\r}$
makes the formal theory invariant under $SO(3)\otimes SO(D-3),$
rather than $SO(D).$ As a result, the Feynman rules
involve hatted and twiddled objects in a non-trivial
way. For the gauge propagator  we have (dropping colour indices):
$
D_{\mn}(\tp,\hp) = \dd_{\mn}(p) +  R_{\mn}(\tp,\hp) ~ ,
$
where $\dd_{\mn}(p)$ and $R_{\mn}(\tp,\hp)$ are given by
$$
\dd_{\mn}(p) = \,\, \sk {m \over p^2\,(p^2+m^2) }\,\,
   \left( m\, \ee_{\m\r\n}\,p^\r
          + p^2 g_{\mn} - p_\m p_\n \right) ~ ,
$$
$$
\eqalign{
R_{\mn}(\tp,\hp) =  \, \sk {m^3 \over p^2\,[(p^2)^2+m^2\,\tp^2]}
    \, \biggl[ \,\,& {\hp^2 \over p^2+m^2}\,
          \left( \ee_{\m\r\n}\,p^\r + p^2 g_{\mn} +
                {m^2 \over p^2}\, p_\m p_\n \right) \cr
       & +  \tp^2 \hat g_{\mn} + \hp_\m \hp_\n -
              p_\m\hp_\n - \hp_\m p_\n  \,\, \biggr] ~. \cr }
$$
Notice that $R_{\mn}(\tp,\hp)$ is vanishing in the limit $D\to3$.
It is obvious that
vanishing objects do not contribute at the tree level.  At higher
perturbative orders however, they could combine with divergent terms
and yield finite contributions.  A careful study [\GMR] of the
convergence properties of 1PI Green functions having at least one
$R_{\mn}(\tp,\hp)$ insertion leads to the conclusion that they are
finite and hence vanishing at $D=3$.  Therefore we can use
$\dd_{\mn}(p)$ as the ``effective'' free gauge field propagator.

The determination of $\gmb_2^{\rm loc}$ necessitates of the knowledge of
three independent Green functions. It is wise to choose the simplest
ones, namely the vacuum polarization, the ghost self-energy and the
$Hcc$ vertex.  The explicit computation of these Green functions presents,
from the point of view of the algebra involved, the same degree of
complexity as QCD.  The situation is much worse from the point of view
of integration, since now we are faced with massive denominators.  The
key observation that makes the computation feasible is that we are
only interested in the asymptotic behaviour of the integrals for large
values of $m$.  We will then use two theorems theorems that enable us
to tell if an integral is vanishing in the limit $m\to\infty$ or not.

A general integral
from a $L$-loop Feynman diagram  is of the form
$$
I(p,m) = \,m^{\beta}\,\int dk \,\,  {{M(k)} \over
          {\prod_i (l_i^2)^{n_i}
            \prod_j (l_j^2 +m^2)^{n_j}}} \,\, , \eqn\ipm
$$
where the integration measure is $ dk = d^3\!k_1\cdots d^3\!k_L $,
$\beta$ denotes an arbitrary real number and $n_i$ are
positive integers.
The momenta $l_i$ are linear combinations of internal and external momenta.
The numerator $M(k)$ is a monomial of degree $n_k$
in the components of the vectors $k_1,\ldots , k_L.$
We call $d$ to the mass dimension of $I(p,m)$ and denote by
$\ird_{\,\min}$ the minimum  between  zero and the lowest infrared degree
at zero external momenta of all
the subintegrals of $I(p,m)$, including
$I(p,m)$ itself. Then the following theorem holds:

\proclaim Theorem 1. { {\sl If the  integral $I(p,m)$ is both
UV and IR convergent by power counting at non-exceptional external
momenta, and the mass dimension $d$ and $\ird_{\,\min}$ defined above
satisfy $ d - \ird_{\,\min} < 0$, then $I(p,m)$ vanishes
when $m$ goes to $\infty.$} }

To formulate the second vanishing theorem we introduce the notation
$[n]=0$ for $n$ even and $[n]=1$ for $n$ odd.
The theorem  states then:

\proclaim Theorem 2. {{\sl If the  integral $I(p,m)$ in \ipm\
is absolutely convergent at zero external momenta
and its mass dimension $d$ satisfies $[n_k] > d $,
then $I(p,m)\to 0$ as $m\to \infty.$} }

Concerning the hypothesis of UV convergence in these two theorems, we
must mention that it can in fact be relaxed [\GMR] for the case of
TMYM, at least up to two loops, to the much weaker condition that the
integral \ipm\ be finite in dimensional regularization.

In the limits $D\to3$ and $m\to\infty$, the result we obtained for the
second order coefficients appearing in the two-loop effective action
\awkward\ are the following
$$
\a_2\,=\,{14\over 3}\, {\left(\yy\right)}^{\!2} \quad , \quad
\beta_2\,=\,{{169+L}\over 72}\, {\left( \yy\right)}^{\!2} \quad,\quad
\gamma_2\,=\,0\,\, ,
$$
with $L=528 \,\ln 2 - 567 \,\ln 3$. Plugging these values into the
expression of $\tilde\a_2$ and taking into account the values of the
one-loop parameters  in the table, we conclude that
$\tilde\a_2$ is zero and therefore the two-loop corrections do not
have observable consequences. Stated differently, there is no two-loop
shift of the parameter $k$ owing to the cohomologically trivial
character of the second order corrections.
We conjecture that  this picture holds true also at higher perturbative
corrections, despite an explicit calculation is still lacking.

It would be also nice to have at our disposal the corresponding
two-loop results for the other BRS invariant regularizations listed in
the previous table, so to get a thorough check of the uniqueness of the
quantum theory parameterized in terms of the bare $k$.  Unfortunately,
what we have presented here is, at present,
 the only instance where a two-loop
computation has beed carried out for CST with a manifestely BRS
invariant regularization scheme.

\refout
\bye